\documentclass{emulateapj}
\usepackage{natbib}

\slugcomment{Appearing in the 2013 May issue of the Astronomical Journal.}
\shorttitle{Late-Time Dust Emission from SN 1995N}
\shortauthors{S.~D.~Van Dyk}

\begin{document}

\title{Late-Time Dust Emission from the Type IIn Supernova 1995N}

\author{Schuyler D.~Van Dyk\altaffilmark{1}}

\altaffiltext{1}{Spitzer Science Center/Caltech, Mailcode 220-6,
  Pasadena CA 91125; email: vandyk@ipac.caltech.edu.}

\begin{abstract}
The Type IIn supernovae (SNe IIn) have been found to be associated with significant amounts of 
dust. These core-collapse
events are generally expected to be the final stage in the evolution of highly-massive stars,
either while in an extreme red supergiant phase or during a luminous blue variable phase. Both
evolutionary scenarios involve substantial pre-supernova mass loss.
I have analyzed the SN IIn 1995N in MCG $-$02$-$38$-$017 (Arp 261), 
for which mid-infrared archival 
data obtained with the {\sl Spitzer Space Telescope\/} in 2009 ($\sim$14.7 yr after explosion)
and with the {\sl Wide-field Infrared Survey Explorer\/} ({\sl WISE}) in 2010 ($\sim$15.6--16.0 yr after 
explosion) reveal a luminous ($\sim 2 \times 10^7\ L_{\odot}$)
source detected from 3.4 to 24 $\mu$m. These observations probe the 
circumstellar material, set up by pre-SN mass loss, around the progenitor star and 
indicate the presence of $\sim 0.05$--$0.12\ M_{\odot}$ of pre-existing, cool dust at $\sim 240$~K.
This is at least a factor $\sim$10 lower than the dust mass required to be produced
from SNe at high redshift, but the case of SN 1995N lends further evidence that highly massive stars 
could themselves be important sources of dust.
\end{abstract}


\section{Introduction}

The origin of cosmic dust is still an open question. The dust grains may well have come from
stars, although dust may also form {\it in situ\/} in the interstellar medium \citep{draine09}.
The presence of $\sim 10^8\ M_{\odot}$ in quasi-stellar objects at redshift $z > 6$ 
(cosmic age $\lesssim$ 700 Myr) indicates that dust formation
in these very young systems must have been rapid \citep{bertoldi03}.
Asymptotic giant branch (AGB) stars, evolved from main sequence stars with 
$M \lesssim 8\ M_{\odot}$, 
are most efficient, but can only contribute significantly to the ``dust 
budget'' in a galaxy after $\sim 400$ Myr
\citep{cherchneff11}, and will dominate dust production for galaxy ages $\gtrsim 1$--2 Gyr
\citep{galliano08}.
A faster dust-production channel is massive stars 
($M_{\rm initial} \gtrsim 8\ M_{\odot}$) and their endpoints as 
core-collapse supernovae (CCSNe).
To account for the total dust mass in these early galaxies, $\sim 1\ M_{\odot}$ of dust from each
SN is necessary \citep{todini01,nozawa03}.
The problem is that, for nearby CCSNe, the most common of which are the SNe II-Plateau (II-P),
the dust yields inferred from mid-infrared observations of the SNe II-P with the 
{\sl Spitzer Space Telescope\/} are in the general range of 
$\sim 4 \times 10^{-5}\ M_{\odot}$ \citep{meikle07} to a few
$\times 10^{-4}\ M_{\odot}$ \citep{kotak09,andrews11a}.
One caveat is that much of the dust formed in these SNe could be cold, particularly for older events.
Evidence to this effect exists in the form of the $\sim 0.4$--$0.7\ M_{\odot}$ at $\sim 17$--23 K found 
by \citet{matsuura11} for SN 1987A with the {\sl Herschel Space Observatory}.

The progenitors of SNe II-P are thought to have initial masses 
$M_{\rm initial} \lesssim 20\ M_{\odot}$ 
\citep{smartt09,vandyk12}. Another class of CCSNe, the SNe II-narrow 
\citep[SNe IIn;][]{schlegel96,filippenko97}, likely arise from
progenitor stars that initially were significantly more massive than this. 
For example, the SN IIn 1998S may have come
from a highly-massive red supergiant (RSG), which, prior to explosion, experienced a ``superwind''
\citep{fransson05}, possibly analogous to the very luminous Galactic RSG VY CMa
\citep{smith09,mauerhan12}. 
On the other hand, indications are that the SN IIn 2005gl was the explosion of
a $M_{\rm initial} > 50\ M_{\odot}$ star which was experiencing a luminous blue variable phase at 
the time \citep{galyam07,galyam09}.
The most extreme SNe IIn may be pair-instability events 
\citep[with $M_{\rm initial} \gtrsim 95\ M_{\odot}$; e.g.,][]{woosley07}.
A number of SNe IIn show evidence, via radio, X-ray, and optical observations 
\citep[e.g.,][]{williams02,mili08,chandra12}, for sustained interaction between the SN 
shock and a dense, massive circumstellar medium (CSM) set up through extensive pre-SN mass 
loss \citep[e.g.,][]{kiewe12}.
The potential, therefore, exists for SNe IIn to be associated with more substantial quantities of dust
than SNe II-P.

Searches for and detections of dust from SNe IIn have been conducted, with some promising results
\citep[e.g.,][]{gerardy02,fox09,andrews11b,fox11,tanaka12}.
\citet{fox11}, for instance, found a range of dust masses associated with their sample of SNe IIn, 
from $\sim 1.5 \times 10^{-4}$ to $\sim 0.1\ M_{\odot}$, within an order-of-magnitude of the required
dust mass from SNe in high-redshift galaxies.
The origin of the infrared (IR) dust emission is uncertain: It could arise from circumstellar dust heated
by the forward SN shock \citep{tanaka12} 
or from dust formed in a cold, dense shell (CDS) within the interaction region between the SN shock 
and the CSM \citep[e.g.][]{smith08}.
\citet{fox11} concluded that the IR emission from SNe IIn detected at late times most likely arises from 
pre-existing CSM dust, which is heated by 
radiation emitted from ongoing shock-CSM interaction.
Two regimes of warm dust (in the shock or CDS) and cool dust (in the heated CSM)  
both could exist for some SNe IIn \citep{stritzinger12}.

Here I consider SN 1995N. 
This SN was discovered on 1995 May 5 at an age $\gtrsim$10 months and was spectroscopically 
classified as a SN~IIn \citep{pollas95}.
The SN host galaxy is MCG~$-$02$-$38$-$017 (Arp 261), at a distance of 24 Mpc \citep{fransson02}.
Optical photometry and spectroscopy of the SN were obtained for over a decade by \citet{gruendl02},
\citet{li02}, and \citet{pastorello05,pastorello11}. 
\citet{fransson02} analyzed {\sl Hubble Space Telescope\/} ({\sl HST}) ultraviolet (UV) and
optical spectroscopy.
SN 1995N is also a luminous and long-lived radio \citep{chandra09} and X-ray 
\citep{fox00,chandra05,zampieri05} source.
\citet{pastorello11} noted that the narrow-width ($< 500$ km s$^{-1}$) components of the emission 
lines, observed by \citet{fransson02}, weakened over time and the high-ionization forbidden lines,
seen at early times, disappeared altogether, while
intermediate-width (3000--4000 km s$^{-1}$)
lines of H, [O {\sc{i}}], [O {\sc{ii}}], and [O {\sc{iii}}], which are thought to be
associated with the unshocked ejecta, increased in prominence at late times.

\citet{fransson02}
determined that the velocities and densities measured and inferred from the narrow
lines in the spectra were typical for the CSM of RSGs.
Based on the overall properties for SN 1995N, they speculated that the progenitor
was analogous to the highly-luminous RSG VY CMa and post-RSG 
IRC +10420, both of which have 
superwinds with mass-loss rates of $\dot M \sim 10^{-4}$--$10^{-3}\ M_{\odot}$ yr$^{-1}$ and
initial masses $M_{\rm ini} \gtrsim 30\ M_{\odot}$. Such a high mass-loss rate is necessary to 
account for the dense CSM and strong SN-CSM interaction.
From their near-IR photometry and spectroscopy, 
\citet{gerardy02} found a significant IR excess from SN 1995N at ages $\sim 730$--2493 d, which
they argued could be explained by an IR echo of the initial flash from the SN, although reprocessing
by dust of the X-ray/UV emission from the CSM interaction could feasibly provide 
additional IR emission at late times.

\section{Observations}\label{observations}

The host galaxy of SN 1995N was observed with {\sl Spitzer\/} with the Infrared Array Camera
\citep[IRAC;][]{fazio04} at 3.6, 4.5, 5.8, and 8.0 $\mu$m
and with the Multiband Imaging Photometer for {\sl Spitzer\/} \citep[MIPS;][]{rieke04} at 24 and 70
$\mu$m on 2009 March 19 and 29, respectively, as part of program 
50454 (PI: B.~J.~Smith). 
(UT dates are used throughout this manuscript.)
I obtained the publicly available data from the {\sl Spitzer\/} Heritage Archive.
A source, as shown in Figure~\ref{figspitzer}, is clearly detected
at all observed wavelengths, except at 70 $\mu$m, for which the observations were most likely not 
sensitive enough. 
In fact, as seen in Figure~\ref{figspitzer}b, it is one of the
brightest objects seen in or around the host galaxy in the 24 $\mu$m data. 
Its position in the 3.6 $\mu$m data is 
RA (J2000) = 14$^{\rm h}$ 49$^{\rm m}$ $28{\fs}30$, 
Dec (J2000) = $-10$\arcdeg 10\arcmin $13{\farcs}7$, 
in very good agreement with the radio position \citep{chandra09}, making it highly likely that this
is the old SN. The separation of the SN from the host undoubtedly contributes to the ease of 
detection in these bands.
For the explosion date of 1994 July 4 assumed by
\citet{fransson02}, the age of the SN when the {\sl Spitzer\/} data were
obtained is $\sim$14.7 yr. 
SN 1995N, then, can be included in the ``transitional'' age range for CCSNe,
discussed by \citet{tanaka12}, between younger SNe and SN remnants.

The SN field was also observed on 2010 January 31/February 1 (age $\sim$15.6 yr) 
and July 31/August 1 ($\sim$16.0 yr) by the {\sl Wide-field Infrared Survey Explorer\/}
\citep[{\sl WISE};][]{wright2010} in all four of its cryogenic bands at 3.4, 4.6, 12, and 22 $\mu$m.
I show in Figure~\ref{figspitzer}c the SN as seen 
in the Atlas Image in band W3 (12 $\mu$m). 
I will discuss the photometry from these {\sl WISE\/} data below.

\section{Analysis}\label{analysis}

I used the MOPEX \citep[MOsaicking and Point source EXtraction;][]{makovoz05a,makovoz05b} 
package provided by the {\sl Spitzer\/} Science Center (SSC) to mosaic the 
individual artifact-corrected Basic Calibrated Data (BCDs) and 
produce a single image mosaic in each band (for both IRAC and MIPS, I left the first frame out
of each set of observations when mosaicking, since it often has a far shorter exposure time than
the rest of the BCDs and therefore adds mostly noise to the mosaic). I also applied the array 
location-dependent photometric corrections to the IRAC BCDs within MOPEX.
Flux measurements were obtained
with point-response function (PRF) fitting within APEX (Astronomical Point source 
EXtractor), in multi-frame mode for IRAC and single-frame mode for MIPS.
I used a spatially-variable PRF mapping, 
provided online by the SSC, for the IRAC image analysis. A single SSC-provided PRF
was used to obtain the photometry from the MIPS 24 $\mu$m mosaic.
The PRF fluxes were aperture- and color-corrected following the prescriptions provided by
the online instrument 
handbooks\footnote{For IRAC, http://irsa.ipac.caltech.edu/data/SPITZER/docs/irac/iracinstrumenthandbook/ 
and, for MIPS, http://irsa.ipac.caltech.edu/data/SPITZER/docs/mips/mipsinstrumenthandbook/.}.
The flux densities for SN 1995N in each band are listed in Table~\ref{tabphot} and
are shown in Figure~\ref{figdust}.

One can see in Figure~\ref{figdust} that the data are nominally consistent with a blackbody spectrum 
at dust temperature $T_{\rm dust}=240$ K. 
The blackbody radius is $R \approx 1.6 \times 10^{17}$ cm ($\approx 10700$ AU or 
$\approx 0.05$ pc).
(This holds if the dust is optically thick, whereas I assume, below, that the dust is optically
thin; so, this radius is effectively a lower limit on the actual size of the dust shell; 
e.g., \citeauthor{fox11}~\citeyear{fox11}.)
For the progenitor wind speed of $\sim 50$ km s$^{-1}$ \citep{fransson02}, matter at this radius 
was lost by the star $\sim 1000$ yr prior to explosion.
Assuming the SN shock was expanding at $\sim 2700$ km s$^{-1}$ \citep{zampieri05}
when the {\sl Spitzer\/} data were obtained,
after $\sim 14.7$ yr the shock radius would have been $\sim 1.3 \times 10^{17}$ cm.
The shock therefore appears to be within the blackbody radius, so it is unlikely that the dust was 
freshly formed in the SN ejecta (the dust is also relatively cold, and, if freshly formed, we would 
expect it to be warmer) or within the CDS in the interaction region. 
It is more likely that the emission arises from pre-existing dust in the unshocked CSM.

\bibpunct[; ]{(}{)}{;}{a}{}{;}

Following \citet{dwek83}, I can estimate the radius, $R_v$, 
of the dust-free cavity that the SN evacuated in the dusty CSM by its initial outburst.
I can constrain the maximum blue luminosity from the optical light curves shown in 
\citet{pastorello05}; note, however, that we do not know the actual maximum, since the SN was 
discovered at $\gtrsim 10$ months in age. 
I therefore estimate that the luminosity $L_0 \gtrsim 2.5 \times 10^{40}$ erg
s$^{-1}$, based on the earliest available data in the $U$ and $B$ bands and assuming 
Galactic foreground extinctions from 
\citet[][NED\footnote{NED is the NASA/IPAC Extragalactic Database, http://ned.ipac.caltech.edu/.}]{schlafly11}.
From \citet[][his Equation 8]{dwek83}, assuming a grain emissivity 
$Q_{\nu}=({\lambda}_0/{\lambda})^2$ (appropriate for silicate grains; see below),
${\lambda}_0=0.2$ $\mu$m, and $T_v \sim 1500$ K (i.e., the evaporation temperature for silicates),
I estimate that this radius is $R_v \gtrsim 1.4 \times 10^{16}$ cm, likely well within both the blackbody 
and shock radius in 2009.

Following \citet{hildebrand83} and \citet{bouchet06}, 
the wavelength-dependent flux density for an optically thin point source (an ``idealized dust cloud'') is 

\begin{equation}\label{eq1}
F_{\nu}({\lambda}) = {{M_{\rm dust} \kappa(\lambda) B_{\nu} (\lambda, T_{\rm dust})} \over {D^2}}
\end{equation}

\noindent where $B_{\nu} (\lambda, T_{\rm dust})$ is the Planck function at the dust temperature
$T_{\rm dust}$, $M_{\rm dust}$ is the dust mass, $D$ is the distance to the host galaxy in Mpc, and 
$\kappa(\lambda)$ is the mass absorption coefficient for the dust. This latter term is given by

\begin{equation}\label{eq2}
\kappa(\lambda) = {{3 Q(\lambda)} \over {4 \rho a}}
\end{equation}

\noindent where $Q(\lambda)$ is the wavelength-dependent dust absorption coefficient, $a$ is the
dust grain radius, and $\rho$ is the grain mass density. Here I have assumed the dust models 
from \citet{laor93} for graphite and smoothed UV  ``astronomical silicates,''  with $\rho\approx 2.2$ g 
cm$^{-3}$ \citep{weingartner01} and $\approx 3$ g cm$^{-3}$ \citep{draine84}, respectively.
I have selected a single grain size, $a=0.1$ $\mu$m, although the choice of $a$
has little impact on $\kappa(\lambda)$ in the Rayleigh limit ($a < \lambda$).

One can see in Figure~\ref{figdust} that the silicate dust model spectrum from Equation~\ref{eq1} 
compares reasonably well with the 8 and 24 $\mu$m data points, less so for the shorter-waveband 
data. Dust models that are predominantly silicates have been applied previously
for a number of SNe \citep[e.g.,][]{bouchet06,meikle07,kotak09,tanaka12}.
Additionally, if the SN 1995N progenitor were analogous to IRC +10420, we might expect the CSM
to be composed primarily of silicate dust \citep[e.g.,][]{blocker99}.
The graphite model provides a better comparison at the shortest bands, however, it is a 
comparatively poorer match at 8 and 24 $\mu$m.
Unfortunately, the overall dust spectrum, 
compared to the {\sl Spitzer\/} data, is relatively unconstrained, since no 
{\sl Spitzer\/} observations
were obtained of this field with the Infrared Spectrograph (IRS) in either Peak-Up Imaging mode at 16
$\mu$m or in Staring spectroscopic mode. (The observations that were conducted, although 
obtained during the cryogenic phase of the {\sl Spitzer\/} mission, did not become publicly 
available till well after the cryogen was exhausted, when IRS observations were no longer possible.)
The high 70 $\mu$m upper limit (not shown) provides a poor constraint on the model. 

Although no additional {\sl Spitzer\/} observations were obtained, I consider
the flux densities for the SN from the {\sl WISE\/} All-Sky Data Release in 
all four bands, using the zero points and color corrections provided in the 
{\sl WISE\/} Explanatory Supplement\footnote{http://wise2.ipac.caltech.edu/docs/release/allsky/expsup/.}.
The flux densities are given in Table~\ref{tabphot2} and shown in Figure~\ref{figdust2}.
Interestingly, the {\sl WISE\/} data tend to agree better with the graphite model than with the silicate 
one.
I note, however, as the Explanatory Supplement warns, that the 
flux densities in bands W1 and W2 at these low levels are systematically underestimated. 
Additionally, a measure of the reliability of a {\sl WISE\/} flux extraction
is a comparison of the number of individual frames for which profile fitting for the photometry was
possible, for sources with a signal-to-noise ratio $S/N \ge 3$ on each frame, versus the total 
number of available individual frames. For these observations, the total in
all four bands was 25 frames; the numbers of frames for which profile-fit flux measurements were
possible were 15, 3, 12, and 0, respectively, for W1, W2, W3, and W4 (the ultimate $S/N$ is only
$\approx 5$ in W4).
Therefore, as a result of the lower sensitivity of the {\sl WISE\/} survey data 
compared to the relatively deep, pointed {\sl Spitzer\/} observations,  
the {\sl WISE\/} data for SN 1995N should possess less weight in the overall analysis.

Integrating the silicate model dust spectrum, the total IR luminosity emitted by the dust would then be
$L_{\rm IR} \approx 8.4 \times 10^{40}$ erg s$^{-1}$, or $\approx 2.2 \times 10^7\ L_{\odot}$.
The IR luminosity of the blackbody alone is $\approx 6.0 \times 10^{40}$ erg s$^{-1}$ (the luminosity
of the graphite model is essentially the same as this).
\citet{gerardy02} modeled the near-IR emission from SN 1995N using an IR echo model from a
pre-existing, spherically-symmetric CSM, following the formalism of \citet{emmering88}.  
I show in Figure~\ref{figlc} the luminosity of the SN in 2009, estimated both from the blackbody 
and the silicate dust model, and compare these with an extrapolation in time of the model 
light curve that \citet{gerardy02} 
constructed with the assumption that the input luminosity from the SN is a short flash. 
One can see that  the echo model underpredicts the observations by a factor $\gtrsim 2$. The
echo alone, therefore, cannot account for the observed SN mid-IR luminosity. 
Note, however, that the model light curve was fit by \citet{gerardy02} to an extrapolation of the 
near-IR color,
which represents only the hottest dust. (Mid-IR observations at earlier times, had they been feasible,
would have provided a better constraint on the total IR evolution
and would have sampled dust at a larger range of temperatures.)
Note also that for this light curve model~\citeauthor{gerardy02}~assumed that some pre-SN event had 
created the dust-free cavity in the CSM, rather than the SN itself, and they treated $R_v$ and $T_v$ 
as being independent of each other. However, we would not necessarily 
expect such an event for a progenitor star analogous to the post-RSG IRC +10420.

The luminosity from the echo model potentially can be augmented if I also take into
account the dust emission resulting from the absorption and reprocessing 
of the X-ray and UV radiation arising from the shock-CSM interaction.
\citet{gerardy02} also pointed out that the IR emission from SN 1995N at very late times could be 
better explained by sustained emission from the CSM interaction.
The X-rays from this interaction were first detected on 1996 July 23 \citep{lewin96,fox00,chandra05},
$\sim 750$ d after explosion (it is unknown when this X-ray emission from the interaction began).
The time-averaged X-ray (0.1--2.4 and 0.5--7.0 keV) unabsorbed flux from the light curves in
\citet{chandra05}
corresponds to $L_{\rm X} \sim 8 \times 10^{40}$ erg s$^{-1}$.
No similar UV light curve exists for the SN, however, a partial estimate of the UV luminosity can be 
obtained by integrating the {\sl HST\/} FOS UV spectrum analyzed by
\citet{fransson02} and by assuming an extinction of $A_{\rm UV}\sim 0.7$ mag
\citep[NED;][]{schlafly11}.
This results in $L_{\rm UV} \sim 7 \times 10^{40}$ erg s$^{-1}$.
From \citet[][their Figure 11]{draine84} the Planck-averaged emissivity of silicate grains at 
$T=240$~K is $\sim 0.35$, irrespective of grain size.
Thus, although without constructing a detailed model, I can account for an additional 
$\sim 5 \times 10^{40}$ erg s$^{-1}$ of reprocessed X-ray and UV
flux from the dust. (This estimate does not include the sustained optical emission from
the CSM interaction.) In total, then, it is feasible that  
the very late-time IR luminosity from SN 1995N is a combination of both 
a fading IR echo from the initial SN flash
and dust re-radiation of the emission produced by the long-term interaction of the shock and the 
CSM.

It is not known what was the level of CSM interaction occurring when the {\sl Spitzer\/} data 
were obtained. The epochs of the {\sl Spitzer\/} and {\sl WISE\/} observations were 
well past the last X-ray and radio data 
obtained by \citet{chandra05} and \citet{chandra09}, respectively. Furthermore, 
the last optical spectrum from 2010 shown in \citet{pastorello11} appears to be dominated 
by the intermediate-width H and forbidden O lines, presumably from the unshocked ejecta.

The dust mass that results from the simple silicate model spectrum (Equation~\ref{eq1}) is 
$M_{\rm dust} \approx 0.05\ M_{\odot}$.
The graphite model, which provides an overall poor comparison with the {\sl Spitzer\/} data, 
requires an even larger dust mass of $ \approx 0.12\ M_{\odot}$.
Assuming a gas-to-dust ratio of $\sim 100$ \citep{savage79}, this implies that the total mass in the 
nebula containing the dust is $\sim 5$--$12\ M_{\odot}$. 
The mass-loss rate from the progenitor star was 
likely high, at $\dot M \sim 2\ \times 10^{-4}\ M_{\odot}$ yr$^{-1}$ 
\citep{zampieri05}. Such a mass-loss rate is consistent with that found for IRC +10420
\citep[$\sim 3$--$6  \times 10^{-4}\ M_{\odot}$ yr$^{-1}$; e.g.,][]{humphreys97}.
\citet{zampieri05} concluded that the wind was clumpy up to $\sim 3.5$ yr of the X-ray light curve
for SN 1995N, and that the CSM density distribution must have been smoother farther out (at later 
times in the interaction).
No indication exists, even at later times \citep[e.g.,][]{pastorello11}, for any extinction other than 
that from the Galactic foreground \citep[e.g.,][]{fransson02}. 
So, the assumption of optically thin dust in the nebula is supported.

However, if the SN 1995N progenitor was analogous to IRC +10420, the latter star is currently 
surrounded by an optically thick dust shell \citep[e.g.,][]{blocker99}.
Much of the dust that existed outside of the dust-free cavity in the inner CSM of SN 1995N 
could have been destroyed by the outgoing shock in the assumed
$\sim 10$ months after explosion and before discovery --- at an expansion velocity of 10000 km 
s$^{-1}$ \citep[the highest velocity inferred from the broad emission line component;][]{fransson02}, 
the shock would have expanded to $\sim 2.6 \times 10^{16}$ cm, or $\sim 16$ \% of the blackbody 
radius. The broad component has weakened in more recent spectra \citep{pastorello11}, so the
expansion has decelerated (as the shock moved through the dense CSM). Nonetheless, as 
calculated above, the shock was already at $\gtrsim 80$~\% of the blackbody radius in 2009, 
so a significant fraction of the dust in the CSM possibly has been destroyed. The star IRC +10420 
shows outer, spherical circumstellar shells ejected from the star $\sim$3000 yr ago 
\citep[e.g.,][]{tiffany10} --- if the SN 1995N progenitor was analogous to this star,
then the dust emission from 2009--2010 
may have been from similar shells in the outermost extent of the 
CSM. In summary, this indicates that the total gas and dust mass of the CSM around the progenitor 
initially may have been substantially larger than the estimate I have made, above, based on the dust 
mass implied by the {\sl Spitzer\/} observations.

\section{Conclusions}\label{discussion}

I have detected very late-time ($\sim$14.7--16 yr) emission in the 
mid-IR (3.4--24 $\mu$m) from the SN IIn 1995N, using archival data from {\sl Spitzer} and {\sl WISE}. 
This emission provides an important probe of the circumstellar environment of the progenitor star.
I have applied a simple silicate dust model, which agrees best with the {\sl Spitzer\/}
data, and have found that this model is consistent with a blackbody temperature of 240 K and a
IR luminosity of $L_{\rm IR} \approx 8.4 \times 10^{40}$ erg s$^{-1}$ for the dust.
A graphite model does not compare as well with the {\sl Spitzer\/} observations, although it does
compare better with the less-sensitive {\sl WISE\/} data.
I conclude that the dust emission did not arise from freshly-formed dust in the ejecta or from
dust condensing in the CDS of the shock interaction region.
This emission more likely arose from pre-existing circumstellar dust, heated by 
a combination of the long-lived IR echo from the
initial SN flash and the 
sustained X-ray/UV (and, probably, optical) flux from the SN shock/CSM interaction, e.g., as modeled 
by \citet{gerardy02}. Much of the original dust in the progenitor pre-SN wind has likely been 
destroyed as the shock has expanded through the CSM.
This result, of heated pre-existing dust  in the CSM, 
is consistent with what has been found for the SN IIn 2010jl by \citet{andrews11b} and 
for the Type ``Ibn'' SN 2006jc by \citet{mattila08}.

\citet{fransson02} speculated that the SN 1995N progenitor was a highly-luminous RSG which had
experienced a super-wind, analogous to the extreme Galactic RSG VY CMa and the post-RSG
IRC +10420.
The size of the dust-emitting region (radius $\sim 10000$ AU) inferred for SN 1995N 
is comparable to the largest
extent of the CSM around VY CMa and IRC +10420, and the mass
in the SN 1995N progenitor nebula is comparable to that found for the 
two Galactic stars \citep{smith01,smith09,tiffany10}.
The mid-IR data, therefore, lends credence to this view of the SN progenitor.

The observed late-time dust emission from SN 1995N implies that the dust mass was 
$\sim 0.05$--$0.12\ M_{\odot}$.
I note that, although the observed dust is relatively cool, $\sim 240$~K, even colder dust, at 
$\sim 20$~K, may exist in even larger quantities and simply was not detected in the mid-IR by 
{\sl Spitzer\/} and {\sl WISE}, as has been found 
for SN 1987A \citep{matsuura11} and, e.g., the Crab Nebula \citep{gomez12}.
Unfortunately, no {\sl Herschel\/} observations of SN 1995N will have been conducted during that
mission's operations.
As has been suggested before, e.g., by \citet{mattila08}, 
evidence is continuing to accrue that the
highly-massive progenitors of these strongly-interacting SNe, rather than the SNe themselves, may
contribute significant amounts of dust to galaxies. Such a scenario could be extendable to 
galaxies early in cosmic history.

\acknowledgements

I thank the referee for a helpful suggestion which improved this manuscript. 
This work is based on observations made with the {\it Spitzer Space Telescope}, which is operated by 
the Jet Propulsion Laboratory, California Institute of Technology under a contract with NASA.
This work also made use of data products from the {\sl Wide-field Infrared Survey Explorer}, which is a 
joint project of the University of California, Los Angeles, and the Jet Propulsion Laboratory/California 
Institute of Technology, funded by NASA.

\clearpage

\begin{deluxetable}{cccc}
\tablewidth{4.7truein}
\tablecolumns{4}
\tablecaption{{\sl Spitzer\/} Mid-Infrared Flux Densities for SN 1995N\label{tabphot}}
\tablehead{
\colhead{Wavelength} & \colhead{Measured} & \colhead{PRF-Corrected\tablenotemark{a}} & 
\colhead{Color-Corrected\tablenotemark{b}} \\
\colhead{} & \colhead{Flux Density} & \colhead{Flux Density} & 
\colhead{Flux Density} \\
\colhead{($\mu$m)} & \colhead{($\mu$Jy)} & \colhead{($\mu$Jy)} & 
\colhead{($\mu$Jy)}
}
\startdata
 3.6 &  $62.6 \pm 1.1$ & 61.3 & 43.8 \\
 4.5 & $124.6 \pm 1.7$ & 123.1 & 100.4 \\
 5.8 & $216.1 \pm 7.1$ & 211.4 & 187.1 \\
 8.0 & $615.9 \pm 8.9$ & 607.4 & 559.0 \\
 24 & $3499.0 \pm 36.3$ & $\cdots$ & 3631.6 \\ 
 70 & $\lesssim$30000 & $\cdots$ & $\cdots$ \\
\enddata
\tablenotetext{a}{Aperture and pixel-phase corrections from the IRAC Instrument Handbook.}
\tablenotetext{b}{Color corrections from the IRAC Instrument Handbook, assuming 
a blackbody at 240 K.}
\end{deluxetable}


\begin{deluxetable}{ccc}
\tablewidth{3.65truein}
\tablecolumns{4}
\tablecaption{{\sl WISE\/} Mid-Infrared Flux Densities for SN 1995N\label{tabphot2}}
\tablehead{
\colhead{Wavelength} & \colhead{All-Sky Release} &  
\colhead{Color-Corrected\tablenotemark{a}} \\
\colhead{} & \colhead{Flux Density} & \colhead{Flux Density} \\
\colhead{($\mu$m)} &  \colhead{($\mu$Jy)} & \colhead{($\mu$Jy)}
}
\startdata
 3.4 &    $91 \pm 6$    &       53 \\
 4.6 &  $103 \pm 12$ &       84 \\
 12 & $1451 \pm 105$ &  1543 \\
 22 & $3541 \pm 760$ &  3595 \\ 
\enddata
\tablenotetext{a}{Color corrections from the {\sl WISE\/} Explanatory Supplement, assuming 
a blackbody at 240 K.}
\end{deluxetable}

\clearpage

\begin{figure}
\figurenum{1}
\includegraphics[angle=0,scale=0.70]{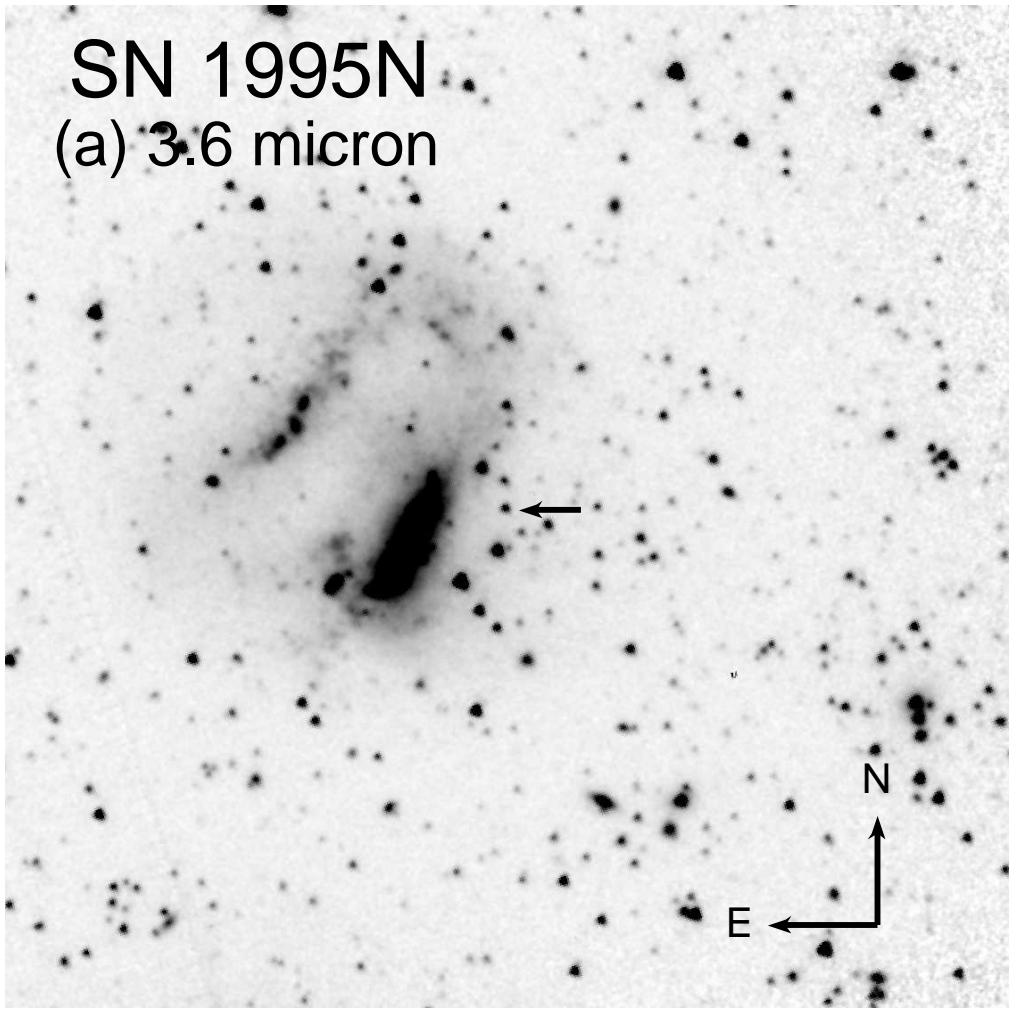}
\caption{Image mosaics at (a) 3.6 $\mu$m and (b) 24 $\mu$m, constructed from observations 
obtained with {\sl Spitzer\/} in 2009 March, of the host galaxy, MCG $-$02$-$38$-$017 (Arp 261),
of the SN IIn 1995N (indicated by the arrow).
The publicly-available data were obtained from the {\sl Spitzer\/} Heritage Archive. 
Also shown in (c) is the {\sl WISE\/} Atlas Image showing the SN
in band W3 (12 $\mu$m), produced from survey observations in 2010.
North is up, and east is to the left.
This figure can be compared with, e.g., \citet[][their Figure 1]{fransson02}.\label{figspitzer}}
\end{figure}

\clearpage

\begin{figure}
\figurenum{1}
\includegraphics[angle=0,scale=0.70]{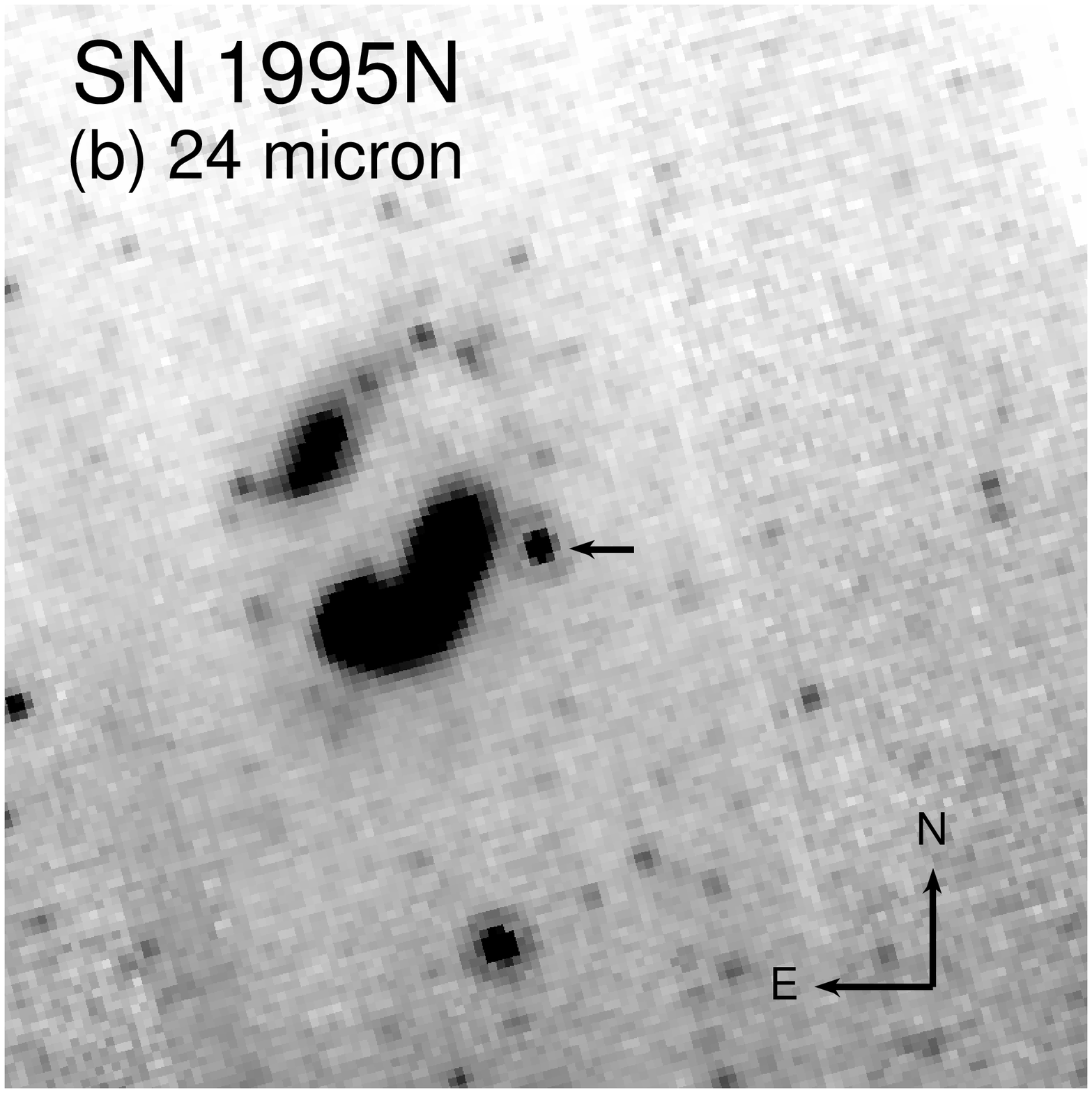}
\caption{(Continued.)}
\end{figure}

\clearpage

\begin{figure}
\figurenum{1}
\includegraphics[angle=0,scale=0.70]{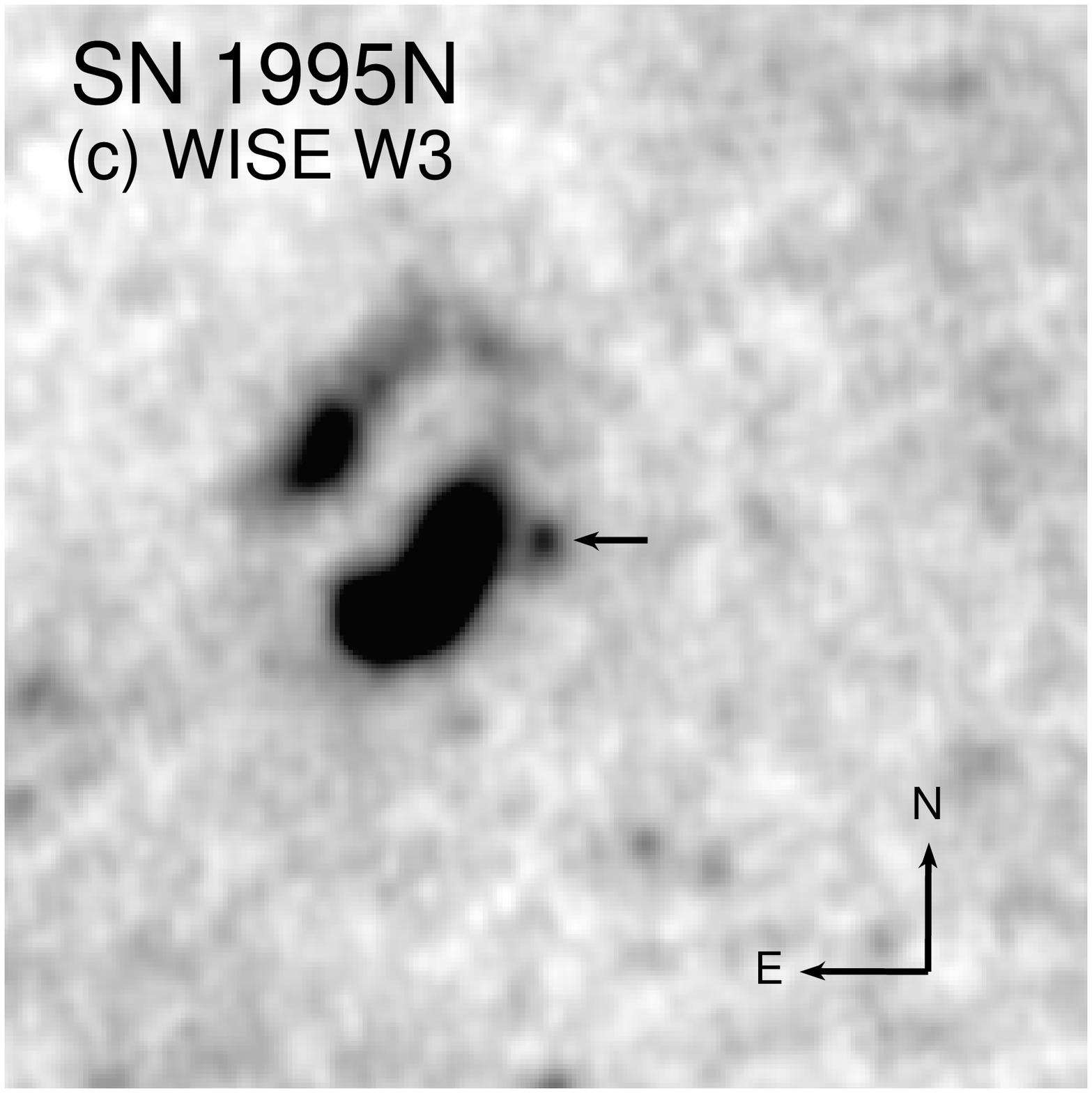}
\caption{(Continued.)}
\end{figure}

\clearpage

\begin{figure}
\figurenum{2}
\includegraphics[angle=0,scale=0.70]{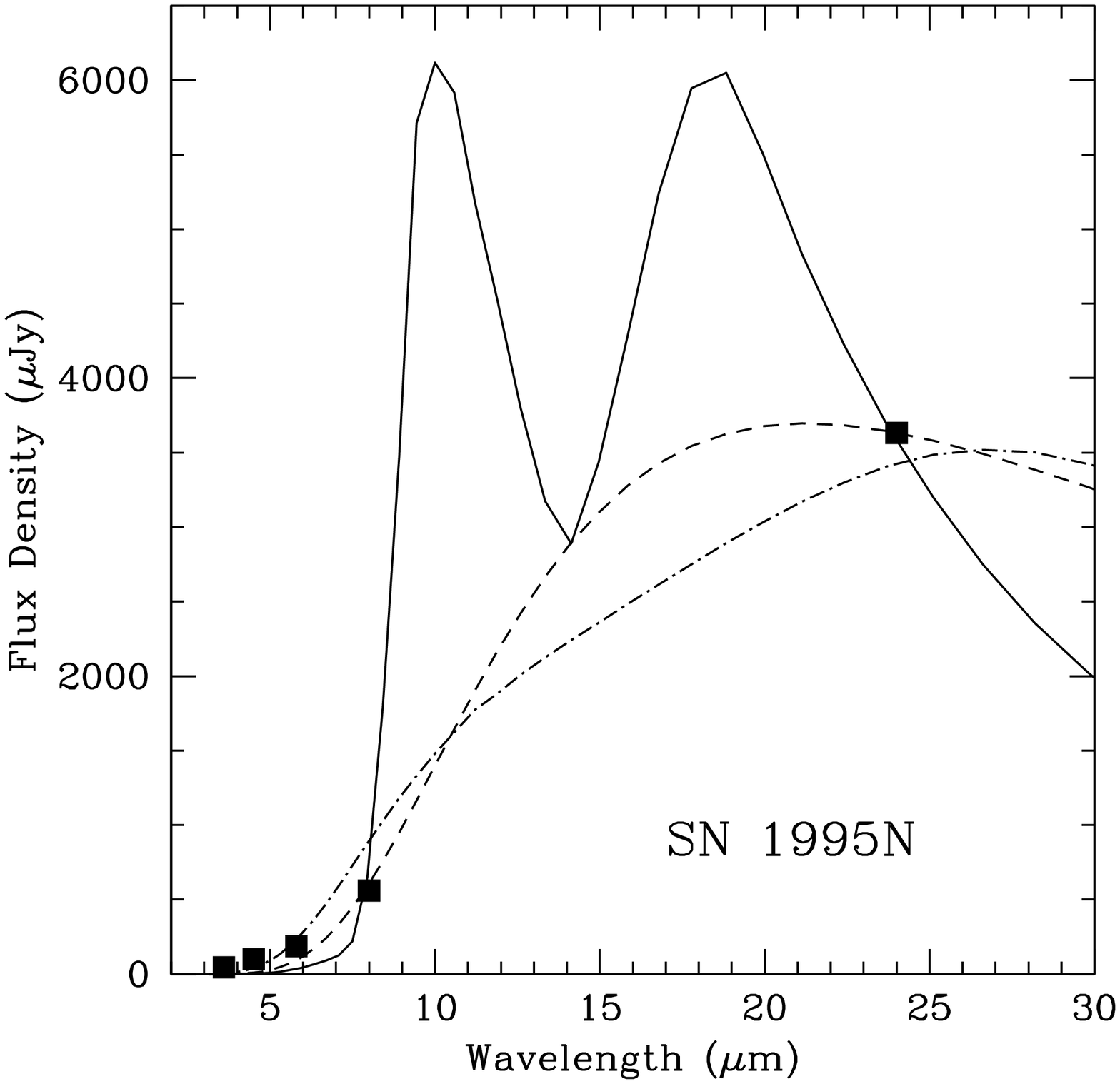}
\caption{Flux densities for SN 1995N ({\it solid points}) measured from the 2009 {\sl Spitzer\/} 
observations of the host galaxy. See Table~\ref{tabphot}. The uncertainties in these measurements
are much smaller than the size of the points shown.
Shown for comparison are a blackbody spectrum at dust temperature $T_{\rm dust}= 240$ K
({\it dashed line}) and model spectra, assuming dust \citep{laor93} composed of 
graphite ({\it dot-dashed line}) and ``astronomical silicates'' ({\it solid line}) at this temperature, as 
described in the text.\label{figdust}}
\end{figure}

\clearpage

\begin{figure}
\figurenum{3}
\includegraphics[angle=0,scale=0.70]{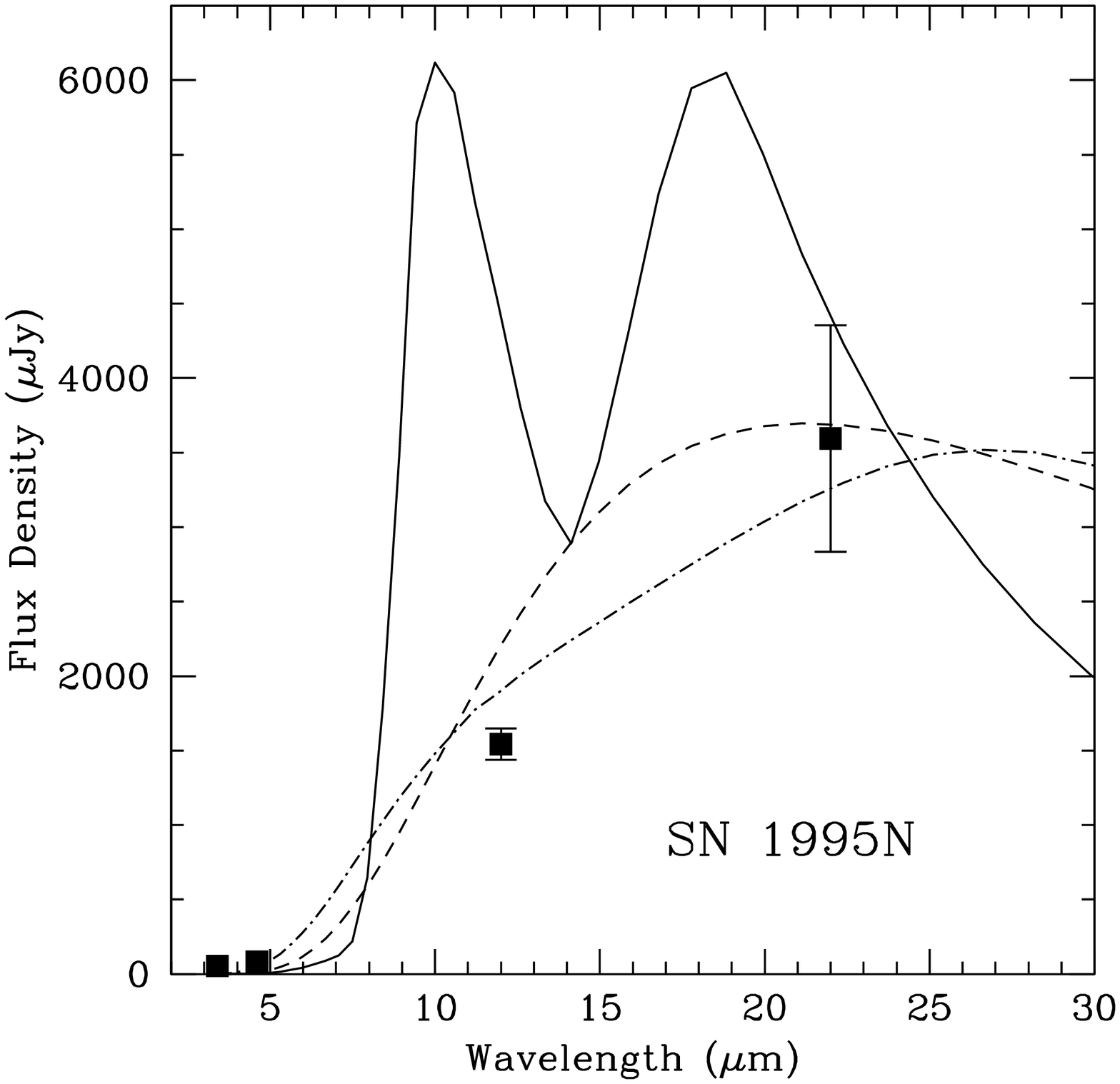}
\caption{Same as Figure~\ref{figdust}, but with flux densities for SN 1995N 
from the {\sl WISE\/} All-Sky Data Release (observations from 2010). See Table~\ref{tabphot2}.
The uncertainties in the flux densities at 3.4 and 4.6 $\mu$m are much smaller than the
size of the points shown. \label{figdust2}}
\end{figure}

\clearpage

\begin{figure}
\figurenum{4}
\includegraphics[angle=0,scale=0.70]{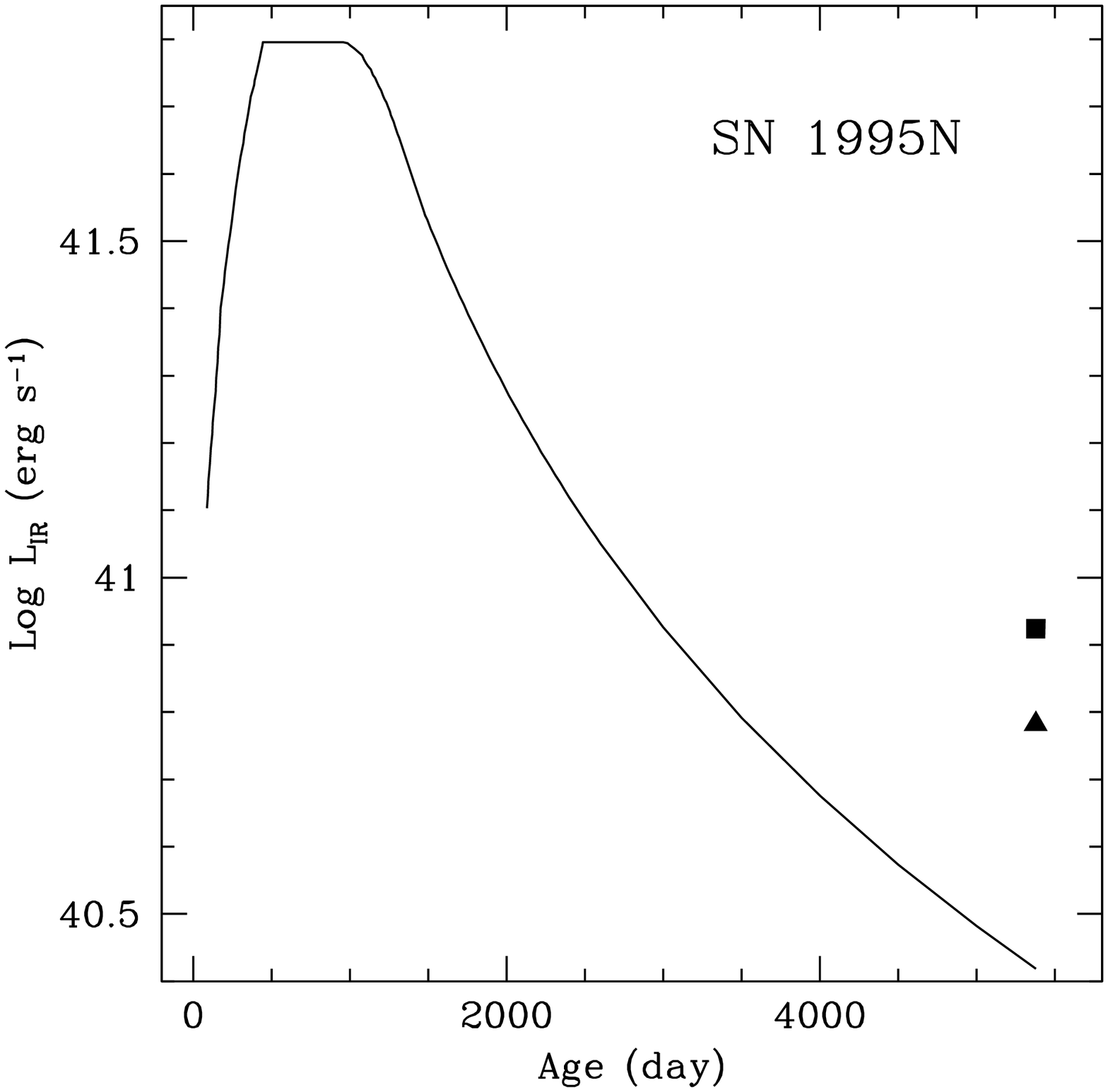}
\caption{The total luminosity from the 2009 {\sl Spitzer\/} observations in the mid-IR of SN 1995N, 
assuming the blackbody at dust temperature 
$T_{\rm dust}=240$ K ({\it solid triangle}) and the silicate dust model ({\it solid square}). See
Figure~\ref{figdust}. Shown for comparison is an extrapolation of the IR light curve ({\it solid line})
constructed by \citet{gerardy02} to fit the early-time, near-IR measurements of the SN, 
following \citet{emmering88} and assuming that the SN was a short flash. Note that this light curve
model underestimates the 2009 mid-IR luminosity by a factor $\gtrsim 2$. The balance of the 
emission plausibly can be provided via reprocessing by the CSM dust of the X-ray/UV flux from the 
long-lived SN shock-CSM interaction.\label{figlc}}
\end{figure}


\begin{thebibliography}{}
\bibitem[Andrews et al.(2011a)]{andrews11a} Andrews, J.~E., 
Sugerman, B.~E.~K., Clayton, G.~C., et al.\ 2011a, \apj, 731, 47
\bibitem[Andrews et al.(2011b)]{andrews11b} Andrews, J.~E., 
Clayton, G.~C., Wesson, R., et al.\ 2011b, \aj, 142, 45
\bibitem[Bertoldi et al.(2003)]{bertoldi03} Bertoldi, F., Carilli, C.~L., Cox, P., et al.\ 2003, \aap, 406, L55 
\bibitem[Bl{\"o}cker et 
al.(1999)]{blocker99} Bl{\"o}cker, T., Balega, Y., Hofmann, K.-H., et al.\ 1999, \aap, 348, 805
\bibitem[Bouchet et al.(2006)]{bouchet06} Bouchet, P., Dwek, E., 
Danziger, J., et al. 2006, \apj, 650, 212
\bibitem[Chandra et al.(2005)]{chandra05} Chandra, P., Ray, A., 
Schlegel, E.~M., Sutaria, F.~K., \& Pietsch, W.\ 2005, \apj, 629, 933
\bibitem[Chandra et al.(2009)]{chandra09} Chandra, P., Stockdale, 
C.~J., Chevalier, R.~A., et al.\ 2009, \apj, 690, 1839
\bibitem[Chandra et al.(2012)]{chandra12} Chandra, P., Chevalier, 
R.~A., Chugai, N., et al.\ 2012, \apj, 755, 110
\bibitem[Draine(2009)]{draine09} Draine, B.~T.\ 2009, Cosmic 
Dust - Near and Far, 414, 453
\bibitem[Draine \& Lee(1984)]{draine84} Draine, B.~T., \& Lee, H.~M.\ 1984, \apj, 285, 89
\bibitem[Dwek(1983)]{dwek83} Dwek, E.\ 1983, \apj, 274, 175
\bibitem[Dwek \& Cherchneff(2011)]{cherchneff11} Dwek, E., \& Cherchneff, I.\ 2011, \apj, 727, 63
\bibitem[Emmering \& Chevalier(1988)]{emmering88} Emmering, R.~T., \& Chevalier, R.~A.\ 1988, 
\aj, 95, 152
\bibitem[Fazio et al.(2004)]{fazio04} Fazio, G.~G., Hora, 
J.~L., Allen, L.~E., et al.\ 2004, \apjs, 154, 10
\bibitem[Filippenko(1997)]{filippenko97} Filippenko, A.~V.\ 1997, \araa, 35, 309
\bibitem[Fox et al.(2000)]{fox00} Fox, D.~W., Lewin, 
W.~H.~G., Fabian, A., et al.\ 2000, \mnras, 319, 1154
\bibitem[Fox et al.(2009)]{fox09} Fox, O., Skrutskie, M.~F., 
Chevalier, R.~A., et al.\ 2009, \apj, 691, 650 
\bibitem[Fox et al.(2011)]{fox11} Fox, O.~D., Chevalier, 
R.~A., Skrutskie, M.~F., et al.\ 2011, \apj, 741, 7
\bibitem[Fransson et al.(2002)]{fransson02} Fransson, C., 
Chevalier, R.~A., Filippenko, A.~V., et al.\ 2002, \apj, 572, 350
\bibitem[Fransson et al.(2005)]{fransson05} Fransson, C., Challis, 
P.~M., Chevalier, R.~A., et al.\ 2005, \apj, 622, 991
\bibitem[Galliano, Dwek, \& Chanial(2008)]{galliano08} Galliano, F., Dwek, 
E., \& Chanial, P.\ 2008, \apj, 672, 214
\bibitem[Gal-Yam et al.(2007)]{galyam07} Gal-Yam, A., Leonard, 
D.~C., Fox, D.~B., et al.\ 2007, \apj, 656, 372
\bibitem[Gal-Yam \& Leonard(2009)]{galyam09} Gal-Yam, A., \& Leonard, D.~C.\ 2009, \nat, 458, 865
\bibitem[Gerardy et al.(2002)]{gerardy02} Gerardy, C.~L., Fesen, 
R.~A., Nomoto, K., et al.\ 2002, \apj, 575, 1007
\bibitem[Gomez et al.(2012)]{gomez12} Gomez, H.~L., Krause, O., 
Barlow, M.~J., et al.\ 2012, \apj, 760, 96
\bibitem[Gruendl et al.(2002)]{gruendl02} Gruendl, R.~A., Chu, 
Y.-H., Van Dyk, S.~D., \& Stockdale, C.~J.\ 2002, \aj, 123, 2847
\bibitem[Hildebrand(1983)]{hildebrand83} Hildebrand, R.~H.\ 1983, 
\qjras, 24, 267
\bibitem[Humphreys et al.(1997)]{humphreys97} Humphreys, R.~M., 
Smith, N., Davidson, K., et al.\ 1997, \aj, 114, 2778
\bibitem[Kiewe et al.(2012)]{kiewe12} Kiewe, M., Gal-Yam, A., 
Arcavi, I., et al.\ 2012, \apj, 744, 10
\bibitem[Kotak et al.(2009)]{kotak09} Kotak, R., Meikle, 
W.~P.~S., Farrah, D., et al.\ 2009, \apj, 704, 306
\bibitem[Laor \& Draine(1993)]{laor93} Laor, A., \& Draine, B.~T., 1993, \apj, 402, 441
\bibitem[Lewin et al.(1996)]{lewin96} Lewin, W.~H.~G., 
Zimmermann, H.-U., \& Aschenbach, B.\ 1996, \iaucirc, 6445, 1
\bibitem[Li et al.(2002)]{li02} Li, W., Filippenko, A.~V., 
Van Dyk, S.~D., et al.\ 2002, \pasp, 114, 403
\bibitem[Makovoz \& Khan(2005)]{makovoz05a} Makovoz, D., \& Khan, I.\ 2005, Astronomical Data 
Analysis Software and Systems XIV, 347, 81
\bibitem[Makovoz \& Marleau(2005)]{makovoz05b} Makovoz, D., \& Marleau, F.~R.\ 2005, \pasp, 
117, 1113
\bibitem[Matsuura et al.(2011)]{matsuura11} Matsuura, M., Dwek, 
E., Meixner, M., et al.\ 2011, Science, 333, 1258
\bibitem[Mattila et al.(2008)]{mattila08} Mattila, S., Meikle, 
W.~P.~S., Lundqvist, P., et al.\ 2008, \mnras, 389, 141
\bibitem[Mauerhan \& Smith(2012)]{mauerhan12} Mauerhan, J., \& Smith, N.\ 2012, \mnras, 424, 2659
\bibitem[Meikle et al.(2007)]{meikle07} Meikle, W.~P.~S., 
Mattila, S., Pastorello, A., et al.\ 2007, \apj, 665, 608
\bibitem[Milisavljevic et al.(2008)]{mili08} Milisavljevic, 
D., Fesen, R.~A., Leibundgut, B., \& Kirshner, R.~P.\ 2008, \apj, 684, 1170
\bibitem[Nozawa et al.(2003)]{nozawa03} Nozawa, T., Kozasa, T., 
Umeda, H., Maeda, K., \& Nomoto, K.\ 2003, \apj, 598, 785
\bibitem[Pastorello et al.(2005)]{pastorello05} Pastorello, A., 
Aretxaga, I., Zampieri, L., Mucciarelli, P., 
\& Benetti, S.\ 2005, 1604-2004: Supernovae as Cosmological Lighthouses, 342, 285 
\bibitem[Pastorello et al.(2011)]{pastorello11} Pastorello, A., 
Benetti, S., Bufano, F., et al.\ 2011, Astronomische Nachrichten, 332, 266
\bibitem[Pollas et al.(1995)]{pollas95} Pollas, C., Albanese, D., Benetti, S., Bouchet, P., \& Schwarz, H. 
1995, IAU Circ., 6170, 1
\bibitem[Rieke et al.(2004)]{rieke04} Rieke, G.~H., Young, 
E.~T., Engelbracht, C.~W., et al.\ 2004, \apjs, 154, 25
\bibitem[Savage \& Mathis(1979)]{savage79} Savage, B.~D., \& Mathis, J.~S.\ 1979, \araa, 17, 73 
\bibitem[Schlafly \& Finkbeiner(2011)]{schlafly11} Schlafly, E.~F., \& Finkbeiner, D.~P.\ 2011, \apj, 
737, 103 
\bibitem[Schlegel(1996)]{schlegel96} Schlegel, E.~M.\ 1996, \aj, 111, 1660
\bibitem[Smartt et al.(2009)]{smartt09} Smartt, S.~J., Eldridge,
J.~J., Crockett, R.~M., \& Maund, J.~R.\ 2009, \mnras, 395, 1409
\bibitem[Smith et al.(2001)]{smith01} Smith, N., Humphreys, 
R.~M., Davidson, K., et al.\ 2001, \aj, 121, 1111
\bibitem[Smith, Foley, \& Filippenko(2008)]{smith08} Smith, N., Foley, R.~J., \& 
Filippenko, A.~V.\ 2008, \apj, 680, 568
\bibitem[Smith, Hinkle, \& Ryde(2009)]{smith09} Smith, N., Hinkle, K.~H., 
\& Ryde, N.\ 2009, \aj, 137, 3558
\bibitem[Stritzinger et al.(2012)]{stritzinger12} Stritzinger, M., 
Taddia, F., Fransson, C., et al.\ 2012, \apj, 756, 173 
\bibitem[Tanaka et al.(2012)]{tanaka12} Tanaka, M., Nozawa, T., 
Sakon, I., et al.\ 2012, \apj, 749, 173
\bibitem[Tiffany et al.(2010)]{tiffany10} Tiffany, C., Humphreys, 
R.~M., Jones, T.~J., \& Davidson, K.\ 2010, \aj, 140, 339
\bibitem[Todini \& Ferrara(2001)]{todini01} Todini, P., \& Ferrara, A.\ 2001, \mnras, 325, 726
\bibitem[Van Dyk et al.(2012)]{vandyk12} Van Dyk, S.~D., Cenko, 
S.~B., Poznanski, D., et al.\ 2012, \apj, 756, 131 
\bibitem[Weingartner \& Draine(2001)]{weingartner01} Weingartner, J.~C., \& Draine, B.~T.\ 2001, 
\apj, 548, 296
\bibitem[Williams et al.(2002)]{williams02} Williams, C.~L., 
Panagia, N., Van Dyk, S.~D., et al.\ 2002, \apj, 581, 396 
\bibitem[Woosley, Blinnikov, \& Heger(2007)]{woosley07} Woosley, S.~E., 
Blinnikov, S., \& Heger, A.\ 2007, \nat, 450, 390
\bibitem[Wright et al.(2010)]{wright2010} Wright, E.~L., 
Eisenhardt, P.~R.~M., Mainzer, A.~K., et al.\ 2010, \aj, 140, 1868
\bibitem[Zampieri et al.(2005)]{zampieri05} Zampieri, L., 
Mucciarelli, P., Pastorello, A., et al.\ 2005, \mnras, 364, 1419
\end{thebibliography}
\end{document}